\documentstyle[aps,eqsecnum]{revtex}

\newcommand{\be}{\begin{eqnarray}}
\newcommand{\ee}{\end{eqnarray}}

\newcommand{\dvec}[2]{{\left( \begin{array}{c} #1  \\
        #2  \\ \end{array} \right)}}

\begin{document}

\title{Finite nuclear size correction to the bound-electron
$g$ factor in a hydrogenlike atom}
\author{D. A. Glazov and V. M. Shabaev}

\address
{Department of Physics, St.Petersburg State University,
Oulianovskaya 1, Petrodvorets, St.Petersburg 198504, Russia}

\maketitle

\begin{abstract}
The finite nuclear size correction to the bound--electron 
$g$ factor in hydrogenlike atoms  is investigated
in the range $Z$=1-20.
An analytical formula for this correction which includes
the non-relativistic and dominant relativistic contributions
is derived. In the case of the $1s$ state, 
the results obtained by this formula are compared with
previous  non-relativistic analytical 
and  relativistic numerical calculations.
\newline
PACS number(s): 31.30. Jv, 31.30. Gs
\end{abstract}


\section{Introduction}

Recent experiments on measuring the bound-electron $g$ factor in 
hydrogenlike carbon
 (${\rm C}^{5+}$) reached an accuracy of about $2\cdot 10^{-9}$
\cite{kn:exp1,kn:exp2}. The same accuracy is expected to be
achieved soon for some other low-$Z$ ions. To obtain the 
corresponding precision in the theoretical predictions for 
the bound-electron $g$ factor,
 the relativistic, QED, nuclear recoil 
and nuclear size corrections 
must be evaluated (see
\cite{kn:Beier2,kn:Beier1,cza01,kn:Karsh_arx} and references therein).
In the present paper
we derive a relativistic formula for the finite nuclear size correction
to the bound-electron $g$ factor in the case of an arbitrary state of
a hydrogenlike atom. This formula provides a sufficiently accurate evaluation
of the correction under consideration in the range $Z=1$--$20$.
In the case of the $1s$ state,
the results obtained by this formula are compared with the results 
obtained by the related non-relativistic formula \cite{kn:Karsh_pl}
and with the results of the relativistic numerical evaluation
\cite{kn:Beier1}.

The relativistic units $(\hbar = c = 1)$ and the Heaviside charge
unit $(\alpha=\frac{e^2}{4\pi},e<0)$ are used in the paper.

\section{Basic formulas}

We consider a hydrogenlike atom, placed in a homogeneous
magnetic field, $\vec{A}(\vec r)=[\vec{B}\times \vec{r}]/2$.
We assume that the interaction of the electron with the
 magnetic field is much smaller than the fine
structure splitting and much larger than the hyperfine
structure splitting, if the nucleus has a nonzero spin.
The energy shift of a level $a$
to the first order in the magnetic field is
\begin{eqnarray}
  \Delta E_a=\langle \Psi_{a} | V_B | \Psi_{a} \rangle,
\end{eqnarray}
where
\begin{eqnarray}
  V_B=\frac{|e|}{2}(\vec B \cdot [\vec r \times \vec\alpha] ).
\end{eqnarray}
Assuming that the vector $\vec B$ is directed along the $z$ axis,
the energy shift reads
\begin{eqnarray}
  \Delta E_a=\mu_0 gBm_j\,,
\end{eqnarray}
where $\mu_0=\frac{|e|}{2m}$ is the Bohr magneton,
 $m_j$ is the $z$-projection of the angular
momentum, and $g$ is the bound-electron $g$ factor. In the case of
 a point-charge
nucleus a simple calculation, based on the Dirac equation, yields 
\cite{kn:Zapr}
\begin{eqnarray}
  g_{\rm D}=\frac{\kappa}{j(j+1)}\left(\kappa 
{{E_{n\kappa}}\over{m}} - \frac12\right).
\label{eq:g_0}
\end{eqnarray}
Here $\kappa=(-1)^{j+l+\frac12}(j+\frac12)$ is the relativistic
angular quantum number, 
$j$ is the total angular
momentum of the electron, $l=j\pm 1/2$ defines the parity of the state,
 $E_{n\kappa}$ is the energy of the state 
\begin{eqnarray}
  E_{n\kappa}=\frac{\gamma + n_r}{N}m\,,
\label{eg:defs}
\end{eqnarray}
$n_r = n - | \kappa |$ is the radial quantum number,
$n$ is the principal quantum number,
$ \gamma = \sqrt {\kappa^2 - (\alpha Z)^2}$, and
$ N = \sqrt {(n_r + \gamma)^2 + (\alpha Z)^2}$.

The finite nuclear size induces a deviation of the $g$ factor from its Dirac
value,
\begin{eqnarray}
g=g_{\rm D}+\Delta g.
\end{eqnarray}
To find the nuclear size correction by perturbation theory, we have
to evaluate the expression
\begin{eqnarray}
  \Delta E_a=2\sum_n^{n \ne a}\frac{\langle \Psi_{a} | V_{\rm F} |
  \Psi_{n}\rangle \langle \Psi_{n} | V_B | \Psi_{a} \rangle}{E_a-E_n}\,,
\end{eqnarray}
where $V_{\rm F}=V-V_{\rm C}$ defines a deviation of the potential from
the pure Coulomb one, $V_{\rm C}=-\alpha Z/r$.
After a simple integration over the angular variables, 
the $\Delta g$ value reads
\begin{eqnarray}
  \Delta g=
  \frac{2\kappa m}{j(j+1)}
  \sum_{n'}^{n'\ne n}
  \frac{\langle n \kappa | V_{\rm F} | n' \kappa \rangle
  \langle n' \kappa | r \sigma_x | n \kappa \rangle}
  {E_{n\kappa}-E_{n'\kappa}}\,,
\label{eg:defd}
\end{eqnarray}
where $| n \kappa \rangle$ is a two-component vector defined by
\begin{eqnarray}
  | n \kappa \rangle = \dvec
           {rg_{n \kappa }(r) }
           {rf_{n \kappa }(r) },
\end{eqnarray}
$g_{n \kappa }$ and $f_{n \kappa }$ are the upper and lower radial
components of the Dirac wave function defined as in \cite{akh},
 $\sigma_x$ is the Pauli matrix
acting in the space of the two-component vectors, and the scalar product
of the two-component vectors is defined by
\begin{eqnarray}
  \langle a | b \rangle = \int\limits_0^{\infty}dr\; r^2 (g_a g_b + f_a f_b)\,.
\end{eqnarray}
The sum (\ref{eg:defd}) can be evaluated analytically using the method of the generalized
virial relations for the Dirac equation proposed in \cite{kn:Shabaev}. Equations
(3.26)--(3.29) of Ref. \cite{kn:Shabaev} can be written as
\begin{eqnarray}
  (E_{n' \kappa}-E_{n \kappa}) \langle n' \kappa | r^s | n \kappa \rangle
  &=& - s \langle n' \kappa | i\sigma_y r^{s-1}  | n \kappa \rangle\,, \\
\nonumber\\
  (E_{n' \kappa}-E_{n \kappa}) \langle n' \kappa | i\sigma_y r^s | n \kappa \rangle
  &=& 2m \langle n' \kappa | \sigma_x r^s  | n \kappa \rangle
  + s \langle n' \kappa | r^{s-1}  | n \kappa \rangle -
\nonumber\\
  & & - 2\kappa \langle n' \kappa | \sigma_z r^{s-1}  | n \kappa \rangle \,,\\
\nonumber\\
  (E_{n' \kappa'}+E_{n \kappa}) \langle n' \kappa | \sigma_z r^s | n \kappa \rangle
  &=& 2m \langle n' \kappa | r^s | n \kappa \rangle
  + s \langle n' \kappa | \sigma_x r^{s-1} | n \kappa \rangle -
\nonumber\\
  & & - 2\alpha Z \langle n' \kappa | \sigma_z r^{s-1}  | n \kappa \rangle \,,\\
\nonumber\\
  (E_{n' \kappa'}+E_{n \kappa}) \langle n' \kappa | \sigma_x r^s | n \kappa \rangle
  &=& - s \langle n' \kappa | \sigma_z r^{s-1} | n \kappa \rangle
  + 2\kappa \langle n' \kappa | r^{s-1}  | n \kappa \rangle -
\nonumber\\
  & & - 2\alpha Z \langle n' \kappa | \sigma_x r^{s-1} | n \kappa \rangle,
\end{eqnarray}
where $\sigma_x$,$\sigma_y$,$\sigma_z$ are the Pauli matrices.
From these equations we obtain
\begin{eqnarray}
  \langle n' \kappa | r \sigma_x | n \kappa \rangle &=&
  \left(\kappa\frac{E_{n\kappa}}{m^2}-\frac{1}{2m}\right) \langle n' \kappa
  | n \kappa \rangle + (E_{n \kappa}-E_{n' \kappa})
\nonumber\\
 && \times \left\langle n' \kappa \left |
  \frac{\kappa}{m^2}\left[\left(E_{n\kappa}-\frac{m}{2\kappa}\right)ri\sigma_y
  +mr\sigma_x+\alpha Z i\sigma_y-\kappa\sigma_z \right]
  \right | n \kappa \right\rangle\,.
\label{eg:c1}
\end{eqnarray}
From equation (\ref{eg:c1}), taking into account that 
\begin{eqnarray}
\langle n' \kappa | n \kappa \rangle =
\delta_{n'n}
\end{eqnarray}
and
\begin{eqnarray}
\sum_{n'}^{n'\ne n}| n' \kappa \rangle \langle n' \kappa | =
  I - | n \kappa \rangle \langle n \kappa |\,,
\end{eqnarray}
 we find
\begin{eqnarray}
  \sum_{n'}^{n'\ne n}\frac{ | n' \kappa \rangle \langle n' \kappa |
  r \sigma_x | n \kappa \rangle }{E_{n \kappa}-E_{n' \kappa}}
  =\left(P-\langle n \kappa |P| n \kappa \rangle \right) | n \kappa \rangle,
\end{eqnarray}
where
\begin{eqnarray}
  P=\frac{\kappa}{m^2}\left[\left(E_{n \kappa}-\frac{m}{2\kappa}\right)ri\sigma_y
  +mr\sigma_x+\alpha Zi\sigma_y-\kappa\sigma_z\right]\,.
\end{eqnarray}
Finally, we should evaluate the expression
\begin{eqnarray}
  \Delta g = \frac{2\kappa m}{j(j+1)}
  \left(\langle n \kappa | V_{\rm F} P | n \kappa \rangle -
  \langle n \kappa | V_{\rm F} | n \kappa \rangle
  \langle n \kappa | P | n \kappa \rangle \right)\,.
\end{eqnarray}
We assume that the nuclear charge distribution is described by a spherically
symmetric density $\rho(\vec r)=\rho(r)$, which is normalized by the equation
\begin{eqnarray}
  \int d{\vec r}\; \rho(\vec r) = 1\,.
\end{eqnarray}
The Poisson equation gives
\begin{eqnarray}
  \Delta V_{\rm F} (\vec r) = 4\pi\alpha Z [\rho(\vec r) - \delta (\vec r)],
\end{eqnarray}
where $\Delta$ is the Laplace operator.
When integrated with $V_{\rm F}$, the
 radial functions $g(r)$ and $f(r)$ can be
approximated by the lowest order term of the expansion in powers of $r$.
It follows we have to evaluate the integral
\begin{eqnarray}
  I=\int_0^{\infty}\limits dr\; r^2 r^{2\gamma-2} V_F\,.
\end{eqnarray}
Using the identity
\begin{eqnarray}
  r^{\beta}=\frac{1}{(\beta + 2)(\beta + 3)}\Delta r^{\beta+2}
\end{eqnarray}
and integrating by parts, we find
\begin{eqnarray}
  I =
  \int_0^{\infty}\limits dr ~ r^2 ~ \frac{1}{2\gamma(2\gamma+1)}
   ~ \Delta r^{2\gamma} ~ V_F=
  \int_0^{\infty}\limits dr ~ r^2 ~ \frac{1}{2\gamma(2\gamma+1)}
   ~ r^{2\gamma} ~ \Delta V_F
\nonumber\\
  = \frac{4\pi\alpha Z}{2\gamma(2\gamma+1)} \int_0^{\infty}\limits dr ~ r^2
  ~ r^{2\gamma} ~ \rho(r)=
  \frac{\alpha Z}{2\gamma(2\gamma+1)} ~ \langle r^{2\gamma} \rangle\,,
\end{eqnarray}
where
\begin{eqnarray}
  \langle r^{2\gamma} \rangle = \int d{\vec r}\; r^{2\gamma} \rho(r)\,.
\end{eqnarray}
For the correction to the $g$ factor we obtain
\begin{eqnarray}
  \Delta g &=& \frac{\kappa^2}{j(j+1)}\cdot
  \frac{\Gamma(2\gamma+1+n_r) 2^{2\gamma-1}}
  {\gamma(2\gamma+1)\Gamma^2(2\gamma+1) n_r!(N-\kappa)N^{2\gamma+2} }
\nonumber\\
&& \times \left\lbrack \left( n_r^2 + (N-\kappa)^2 \right)
  \left( 1 - 2\kappa\frac{E_{n \kappa}}{m} \right) -
  2n_r(N-\kappa) \left( \frac{E_{n \kappa}}{m} - 2\kappa \right) \right\rbrack
\nonumber\\  
&&\times (\alpha Z)^{2\gamma+2} m^{2\gamma} \langle r^{2\gamma} \rangle\,.
\label{eq:d_rel}
\end{eqnarray}
For  $ns$-states, which are of particular interest,
the expansion of this expression to two lowest orders in $\alpha Z$ yields
\begin{eqnarray}
  \Delta g &=&\frac{8}{3n^3} (\alpha Z)^4 m^2 \langle r^2 \rangle
  \Biggl[ 1+(\alpha Z)^2
  \Biggl( \frac{1}{4} + \frac{12n^2-n-9}{4n^2(n+1)} 
\nonumber\\
&& + 2\Psi(3) -\Psi(2+n)
   - \frac{\langle r^2 \log(2\alpha Zmr/n) \rangle}
  {\langle r^2 \rangle} \Biggr) \Biggr]\,,
\label{eq:d_nr1n}
\end{eqnarray}
where $\Psi(x)=\frac{d}{dx}\log\Gamma(x)$.
For the $1s $ state, we have
\begin{eqnarray}
  \Delta g = \frac{8}{3} (\alpha Z)^4 m^2 \langle r^2 \rangle
  \Biggl[ 1+(\alpha Z)^2
  \Biggl(2-C - \frac{\langle r^2 \log(2\alpha Zmr) \rangle}
  {\langle r^2 \rangle} \Biggr) \Biggr]\,,
\label{eq:d_nr1}
\end{eqnarray}
where $C$=0.57721566490 is the Euler constant.
In the non-relativistic limit, we find
\begin{eqnarray}
 \Delta g = \frac{8}{3n^3} (\alpha Z)^4 m^2 \langle r^2 \rangle
\label{eq:d_nr}
\end{eqnarray}
for $ns$ states and
\begin{eqnarray}
  \Delta g = \frac{2(n^2-1)}{3n^5} (\alpha Z)^6 m^2 \langle r^2 \rangle
\end{eqnarray}
for $np_{\frac12}$ states. In the case of the $1s$ state,
the expression (\ref{eq:d_nr})
coincides with the related formula in \cite{kn:Karsh_pl}.

\section{Numerical results}

In Table 1 we compare the $\Delta g$ values for the $1s$ state
 obtained
by formula (\ref{eq:d_nr1}) with
the non-relativistic results of Ref. \cite{kn:Karsh_pl}
(it corresponds to  equation (\ref{eq:d_nr})
of the present paper) and with the relativistic numerical
results of Ref. \cite{kn:Beier1}.
To calculate $\langle r^2 \log{r}\rangle $ in equation (\ref{eq:d_nr1}),
we considered the homogeneously charged sphere model for the nuclear charge
distribution.
 As one can see from the table, the relativistic contribution
to $\Delta g$ 
 becomes comparable with the current experimental
accuracy for ions with $Z \geq 12$.
It will be also important for lower $Z$ ions,
provided the experimental accuracy is improved by an order of
magnitude.

\section*{Acknowledgements}

Valuable conversations with T. Beier, S. Karshenboim,
J. Kluge, W. Quint, and V. Yerokhin are gratefully
acknowledged. This work was supported in part by
RFBR (Grant N. 01-02-17248) and by the program
"Russian Universities - Basic Research"
(project No. 3930).

\newpage

\begin{table}
\caption{The finite nuclear size correction $\Delta g$ for the $1s$ state.}
\vspace{0.5cm}
\begin{tabular}{l l l l  l} \hline
$Z$ & $\langle r^2 \rangle^{1/2}$ & Ref.\cite{kn:Karsh_pl}
(= Eq. (\ref{eq:d_nr}))
& Eq. (\ref{eq:d_nr1}) 
&  Ref.\cite{kn:Beier1} \\
& fm & $[10^{-9}]$ & $[10^{-9}]$  & $[10^{-9}]$ \\
\hline
1  & 0.862     & 0.00003768 & 0.00003770  & $<$ 0.01 \\
2  & 1.671     & 0.0022655  & 0.0022705    & $<$ 0.01 \\
4  & 2.390     & 0.074154   & 0.074741      & 0.09 \\
6  & 2.468     & 0.40031    & 0.40710        & 0.42 \\
8  & 2.693     & 1.5064     & 1.5499          & 1.56 \\
10 & 3.006     & 4.5822     & 4.7810          & 4.78 \\
12 & 3.057     & 9.827      & 10.426          & 10.40 \\
14 & 3.123     & 19.00      & 20.54           & 20.47 \\
16 & 3.263     & 35.38      & 39.05            & 38.90 \\
18 & 3.427     & 62.52      & 70.53            & 70.28 \\
20 & 3.478     & 98.15      & 113.4           & 113.15 \\
\hline
\end{tabular}
\end{table}

\end{document}